# From Perception to Cognition: How Latency Affects Interaction Fluency and Social Presence in VR Conferencing

Jiarun Song, *Member, IEEE*, Ninghao Wan, Fuzheng Yang, *Member, IEEE*, Weisi Lin, *Fellow, IEEE*

***Abstract*—Virtual reality (VR) conferencing has the potential to provide geographically dispersed users with an immersive environment, enabling rich social interactions and user experience using avatars. However, remote communication in VR inevitably introduces end-to-end (E2E) latency, which can significantly impact user experience. To clarify the impact of latency, we conducted subjective experiments to analyze how it influences interaction fluency from the perspective of quality perception and social presence from the perspective of social cognition, comparing VR conferencing with traditional video conferencing (VC). Specifically, interaction fluency emphasizes user perception of interaction pace and responsiveness and is assessed using the Absolute Category Rating (ACR) method. In contrast, social presence focuses on the cognitive understanding of interaction, specifically whether individuals can comprehend the intentions, emotions, and behaviors expressed by others. It is primarily measured using the Networked Minds Social Presence Inventory (NMSPI). Building on this analysis, we further investigate the relationship between interaction fluency and social presence under different latency conditions to clarify the underlying perceptual and cognitive mechanisms. The findings from these subjective tests provide meaningful insights for optimizing the related systems, helping to improve interaction fluency and enhancing social presence in immersive virtual environments.**

## I. INTRODUCTION

With the advancement of immersive media and real-time communication technologies, affordable and high-quality virtual reality (VR) systems have created new opportunities for social VR applications [1]. Platforms such as Horizon Workrooms, Remio VR, and MeetinVR offer 3D user virtual spaces to interact through VR head-mounted displays (HMDs) [2]-[4]. As a potential alternative to traditional video conference (VC), VR conferencing is becoming increasingly

This work was supported in part by the National Natural Science Foundation of China (62171353) *(Corresponding author: Weisi Lin).*

Jiarun Song and Ninghao Wan are with the School of Telecommunications Engineering, Xidian University, Xi'an, 710071, China (e-mail: jrsong@xidian.edu.cn; ninghaow@stu.xidian.edu.cn).

Fuzheng Yang is with the School of Telecommunications Engineering, Xidian University, China, and with the School of Electrical and Computer Engineering, Royal Melbourne Institute of Technology, Melbourne, VIC 3001, Australia (e-mail: fzhyang@mail.xidian.edu.cn).

Weisi Lin is with the College of Computing and Data Science, Nanyang Technological University, Singapore 639798 (e-mail: wslin@ntu.edu.sg).

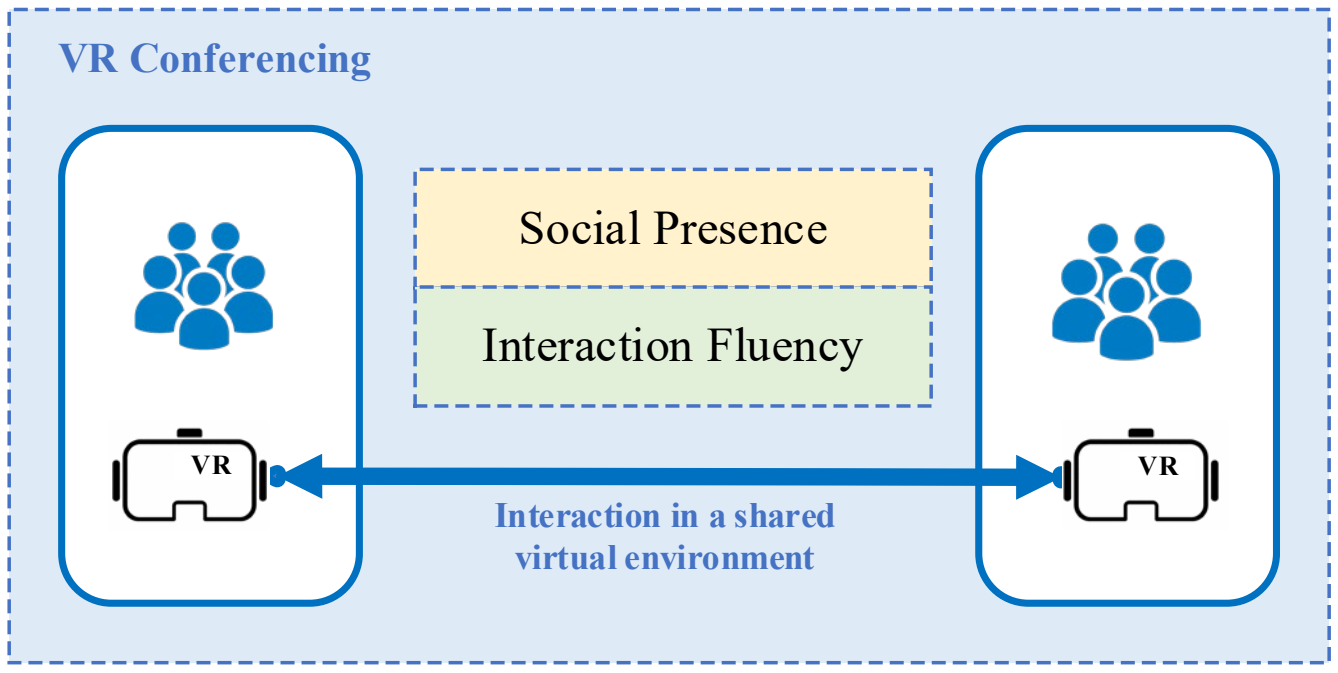

**Fig.1** Framework of VR conferencing.

popular, where users can interact in real time through avatars in a shared virtual environment. Significant efforts have been made to design avatars that closely mimic real-life appearances, gestures, and movements, enhancing realism and enabling richer social interactions. By using these non-verbal and more complex social cues, VR conferencing offers users a higher level of social experience compared to traditional VC platforms [5]-[9].

Interaction fluency and social presence are key metrics for evaluating VR conferencing from both perceptual and cognitive perspectives [10]. Whether in pairwise or multi-user interactions, these two dimensions consistently form the essential foundation of user overall experience in mediated conferencing. As illustrated in Fig. 1, interaction fluency tends to be assessed from the perspective of quality perception, focusing on the user perception of interaction pace and responsiveness [11]. In a conversation, users naturally have expectations about response time. If the perceived response delay exceeds the expected time, the interaction pace will be disrupted, thereby reducing the perceived fluency. For instance, when a user asks a simple question such as "What's the weather like", a significantly delayed response from the other party may disrupt the fluency of the conversation. In this case, users may describe their perception as: "I feel like our conversation is not flowing smoothly". In contrast, social presence is evaluated from a social cognition perspective, emphasizing the user's ability to interpret others' intentions, emotions, and behaviors in a mediated environment [12]. For example, if a user tells a joke but the partner's response is delayed due to latency, the user may question whether the partner understood the joke or is still paying attention. This can also lead to uncertainty about the partner's emotion, with statements such as "I don't think I understand my partner's emotion" (see Appendix: Message Understanding, and Affective Understanding items). Compared to interaction fluency at the perceptual level, the cognitive dimension of social presence covers a wider range of aspects, including the recognition of intention, emotion, and behavior.

Actually, the primary goal of both traditional VC and VR conferencing is to facilitate remote communication between users located in different geographical areas. However, both forms of communication suffer from end-to-end (E2E) latency, which refers to the delay between a user performing an action and others perceiving that action [13], [14]. E2E latency can significantly affect interaction fluency and social presence. Regarding interaction fluency, latency disrupts the natural pace and responsiveness of communication, leading to a sense of discontinuity and reduced coordination during interaction. For social presence, latency affects user's ability to interpret and respond to social cues, thereby impairing their understanding of others' intention and emotion.

While latency in traditional VC has been studied [15]-[19], with a focus on its impact on the overall user experience and mitigation strategies, research on latency in VR conferencing remains comparatively limited, despite emerging efforts in recent years. In fact, the impact of latency may vary significantly across different environments, especially given the unique features of VR, such as wider field of view (FOV) [20], avatar representation, and embodiment, all of which are substantially different from traditional VC settings. In VC, users primarily communicate via real-time video on a PC or mobile device, where interactions are mostly confined to a head-and-shoulder frame within a limited 2D FOV. In contrast, VR communication typically relies on virtual avatars (e.g., Meta Horizon Workrooms), allowing users equipped with HMD to experience a wider FOV. This allows users to see partner's half-/full-body representations and wider surroundings. In addition, VR interactions support 6 degrees-of-freedom (6DoF), which provides a greater sense of immersion. These fundamental differences may lead to different perception and cognition of users in VR conferencing than in VC. Therefore, existing VC research results cannot be directly applied to VR conferencing scenarios without targeted investigations.

Recent studies have started to investigate the perceptual impact of latency in social XR and immersive communication scenarios [21], [22]. In particular, frameworks have been proposed to categorize latency sources into viewport rendering, local interaction, and distant reality delays and analyze the distinct impacts on QoE [21]. Collaborative experiments further quantified acceptable E2E latency thresholds for XR communication [22], while task-oriented VR studies examined the effects of visual and interaction latency on task performance, simulator sickness, and subjective experience [23]. Meanwhile, standardized QoE evaluation methodologies have been established by the International Telecommunication Union (ITU), including interactive testing and latency impact assessment procedures for multiparty communication (e.g., ITU-T P.920 [24], P.1301 [25], and P.1305 [26]), as well as immersive video evaluation guidelines such as P.919 [27], which provide a methodological foundation for XR QoE studies.

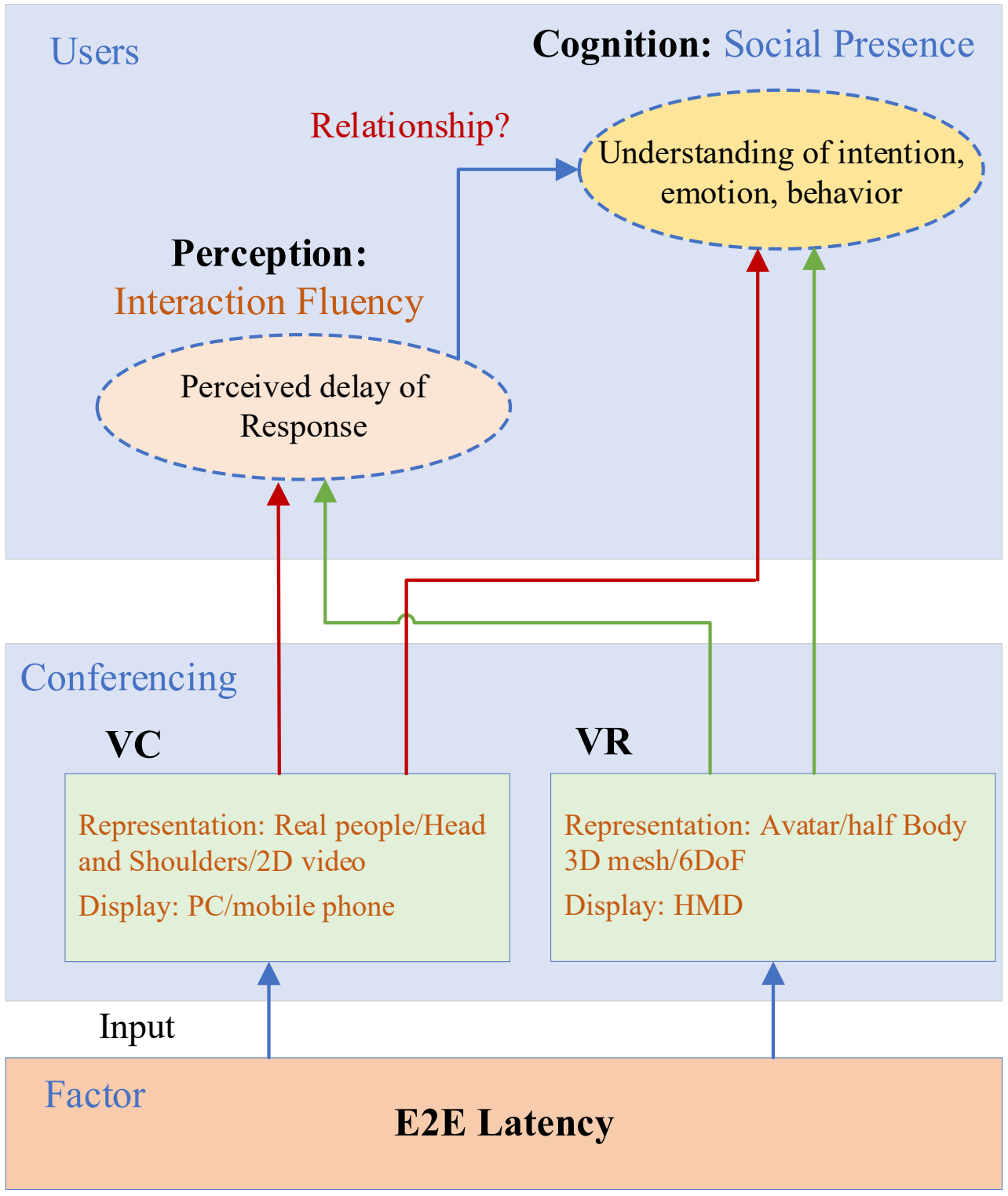

**Fig.2** Research Framework of VR conferencing influenced by E2E latency.

Despite these efforts, existing studies have not thoroughly analyzed the differences in perceptual and cognitive responses to latency across communication modalities. The impact of latency on interaction fluency at the perception level and social presence at the cognition level, as well as their interrelationship, remain insufficiently understood. In VR conferencing with avatar-mediated interaction and VC with real-person video representation, identical levels of perceived interaction fluency under latency may lead to different cognitive evaluations of social presence. For example, when latency exceeds a certain threshold, users participating in the same conversation (e.g., discussing a specific topic) may experience different interpretations depending on the platform. Moreover, these interpretation differences may vary across different latency conditions.

To fill the aforementioned research gap, our previous research specifically examined the impact of latency on social presence in VR conferencing, with a particular emphasis on its impacts on the user's cognitive interpretation of interaction [28]. In the study documented in this paper, as illustrated in Fig. 2, considering the unique characteristics of VR conferencing, such as wider FOV, avatar representation, and embodied interaction, how E2E latency influences user perception of interaction fluency has been first investigated. Furthermore, how perceived interaction fluency affects the user's cognitive processing of interpretation has been explored, and the underlying mechanism has been clarified between perception and cognition. Additionally, the impact of latency in VR conferencing has been compared with traditional VC, to identify the potential reasons contributing to these differences. This comparison provides insights into strategies for optimizing latency in VR conferencing.

The contribution of this study is to address the following exploratory research questions: (1) How does latency affect the perceptual dimension of interaction fluency and the cognitive dimension of social presence during conversations? What is the relationship between interaction fluency and social presence, and what insights can be derived to optimize VR conferencing systems? (2) What are the differential impacts of latency on interaction fluency and social presence between VR and VC conferencing? What are the potential reasons contributing to these differences? By investigating the dual impact of latency on interaction fluency and social presence (from quality perception to social cognition), this research aims to advance immersive VR technology and applications in the metaverse. The findings will provide practical guidelines for improving user experience in VR conferencing.

The remainder of this paper is organized as follows. Section II contains the most related work. Section III discusses the experimental design, including the test platform, participants, and test procedures. Section IV analyzes the experimental results, where the impacts of latency on interaction fluency and social presence are analyzed. Section V comprises the conclusion and further discussion.

## II. Related Work

### *A. Interaction fluency*

Hornbæk [29] provides a comprehensive review of interaction theories and concepts, highlighting various interpretations of the human-computer relationship and their implications for design. A common understanding of interaction is that it occurs only when at least two participants are engaged in a communicative activity. The nature of these participants and the way they interact are key distinctions in existing conceptions of interactivity. Stromer et al. [30] divide interaction into two categories: (1) interaction as a process, and (2) interaction as a product. Interaction as a process focuses on the exchange between participants, where responses are coherent reactions to previous messages or requests, and the roles of participants can be interchangeable. Interaction as a product refers to the set of technical features that enable users to interact with the system. Viewing interaction as a process is more appropriate in the context of multimedia research. For VR conferencing, we adopt Egger's [11] definition of interaction as a sequence of actions, references, and responses, where each action or response is temporally and contextually related to previous events in a predictable and recognizable way.

Interaction fluency mainly reflects the user's perception of the pace and responsiveness of the interaction. Usually, the interaction follows an inherent pace where users can promptly respond to messages or inputs from other participants, thereby minimizing the effort required to maintain seamless communication. However, E2E latency can disrupt this natural pace, resulting in perceived discontinuity and loss of fluency in the interaction. There is substantial research on the impact of latency on interaction fluency. For instance, [31] focused on multiplayer VR games and found that latency exceeding 100 milliseconds (ms) can seriously affect the fluency and responsiveness of the interaction. In related research, Vlahovic et al. [32] studied the impact of network latency on user experience in first-person shooter VR multiplayer games. Their results showed that if the latency exceeds approximately 100ms (round-trip time between client and server), the user's QoE begins to be affected. Kojic et al. [33] explored latency in VR exergaming, observing that users felt a significant decrease in fluency when the delay was set to 500ms.

While the effect of latency on gaming has been well studied, its impact on VR conferencing environments remains relatively unexplored. Unlike the high-intensity, goal-driven interactions characteristic of gaming environments, VR conferencing primarily facilitates Face-to-Face (FTF) communication, emphasizing natural conversations and interpersonal exchanges. Therefore, it is crucial to investigate how latency affects the fluency of interaction in these settings. Understanding this difference can help optimize VR conferencing systems, where the pace of communication may be less critical than in fast-paced VR games.

### *B. Social presence*

Social presence was first conceptualized by Short et al. as the salience of interactants and their interpersonal relationship during mediated communication [34]. According to Short's theory of social presence, intimacy and immediacy are its two core components. Intimacy describes the degree of closeness during interaction, while immediacy involves the psychological distance between the interactants. These two concepts are closely related and are influenced by both verbal and non-verbal cues such as facial expressions, tone of voice, gestures, body posture, and eye contact. Since different media vary in their ability to transmit these social cues, the level of social presence and interpersonal salience experienced by users depends significantly on the media used.

FTF interaction is the ideal benchmark for evaluating the effectiveness of mediated communication technologies in supporting interaction [35]. The concept of social presence emerged from research into this non-mediated form of communication, where both intimacy and immediacy are naturally fostered. While current mediated forms of communication cannot fully replicate FTF interaction, they remain essential substitutes [36]. Biocca et al. define social presence as the degree to which users understand the intelligence, intention, or emotion of others in a mediated environment [37]. The core concepts include two dimensions: co-presence and social connection. Co-presence is the basic sense of being together with others in a virtual space, it involves not only the cognition that others are sharing the same virtual environment, but also the establishment of mutual awareness [38]. In other words, users can be aware of each other's presence and achieve mutual response in interaction. Social connection requires interaction and engagement with others, with a focus on understanding other's intentions, emotions, and behaviors and exerting mutual influence to make them interdependent throughout the interaction process. This perspective shifts social presence from the mere perception of FTF communication to the cognitive process of another intelligent entity.

The factors influencing social presence can be generally classified into two categories: technical factors and human factors. Technical factors include modality, visual representation, display characteristics, and other elements, while human factors include demographic characteristics and psychological traits. For the technical factors, early research often focused on how different degrees of immersion provided by various modalities influenced social presence. Comparisons have been made between computer-mediated communication (CMC) and FTF interactions, audiovisual modalities versus text-based CMC, and immersive versus non-immersive virtual environments [39]-[41]. Moreover, the visual representation of the communication partners was also a focus, especially the impact of the presence or absence of visual representation and the realism of virtual representations on social presence [42], [43]. Additionally, display characteristics, such as screen resolution and size, have been explored for their influence on social presence [44]. Moreover, Jin [45] found that participants experienced greater social presence when their partner's virtual avatar closely resembled their partner's physical appearance. Regarding agency, studies suggest that social presence is higher when users believe that the virtual entity is controlled by a real person rather than a computer program [46]. For the human factors, researchers have also studied individual differences, including psychological characteristics and demographic factors like gender and age [47], [48]. Research indicates that women experience higher levels of social presence than men, and age appears to have little correlation with social presence.

In VR conferencing environments, since participants communicate and interact in different locations, this process will inevitably produce latency. In fact, latency may disrupt the natural flow of interaction and the cognitive process of being "together" with others, thereby significantly affecting social presence. Consequently, it is crucial to understand the impact of latency on social presence. Although several studies have compared traditional VC systems with VR systems [49]-[53], most have focused on the effects of factors such as modality, display, visual representation, and agency. However, to the best of our knowledge, there is still a significant gap in research that specifically examines the impact of E2E latency on social presence in VR conferencing. The relationship between interaction fluency and social presence remains unclear. In this regard, this study focuses on examining the influence of latency on social presence in VR conferencing. By designing a VR conferencing system, the impact of other factors except latency can be kept as constant as possible through the method of single variable control.

## III. Design of Experiments

In order to investigate the impact of E2E latency on interaction fluency and social presence when people interact in VC and VR conferencing, respectively, two subjective experiments were designed and conducted. This section describes the experimental platforms, participants, and test procedures in detail.

### *A. Experimental platforms*

Two custom applications for VR conferencing and traditional PC-based VC were developed, as shown in Fig. 3. The VR conferencing test platform was built on Meta Quest 2 using the Unity Engine [54]. A 3D virtual conferencing space inspired by Horizon Workrooms [55] was developed, where users can interact with others in a VR conferencing environment, and rate their experience on the interaction fluency and social presence. The avatars were created using the Meta Avatars SDK [56] to facilitate interaction, where hand tracking, facial tracking, body tracking, and audio input were used to animate the expressive avatars, ensuring realistic body positions and behavioral realism. The Meta Quest Pro headset, which supports facial expression capture, was used to collect motion, facial, and audio data. In order to be as close as possible to the traditional representation of users in a PC-based video conferencing, the avatars were only represented with the upper body in the VR experiments. The use of an upper body representation instead of a full body representation seems not to affect the social presence when avatars are sitting behind a table [57]. Avatars were first placed close to each other. Then, users could adjust their positions as needed at the beginning of the subjective task. Regarding the traditional VC environment, a PC-based application was developed as a subjective experimental test platform for VC on the Windows system. The platform interface was designed with reference to conferencing systems like Tencent Meeting and Zoom.

In both conferencing applications, transmission was implemented using Web Real-Time Communication (WebRTC) [58] technology, a widely adopted technology for real-time audio-video communication. All experiments were conducted in a controlled local area network (LAN) environment with high and stable bandwidth to minimize network-induced visual quality fluctuations such as compression degradation or stalling. The platform configurations were fixed without manipulating resolution, bitrate, or encoding parameters, thereby keeping image quality as constant and high as practically possible throughout the experiments. The E2E latency was precisely controlled by adjusting the rendering time and calculated as the difference between the Network Time Protocol (NTP) timestamp of the capture time at the sender and the rendering time at the receiver. It includes the latency introduced by all processing stages, i.e., capture, serialization, transmission, deserialization, and rendering. Based on the measurements on five popular social VR platforms [14] and traditional WebRTC-based video transmission [59], the typical E2E latency currently achievable is around 100~150ms. This value will of course increase as users are further away from each other and the number of participants increases [14], [59]. In our setting, the inherent E2E latency was about 80ms. Previous studies have shown that conversations are negatively impacted by E2E latency exceeding 500ms [60]. Taking the above factors into account, we set the E2E latency range to 100-3000ms by adjusting the intrinsic system latency accordingly.

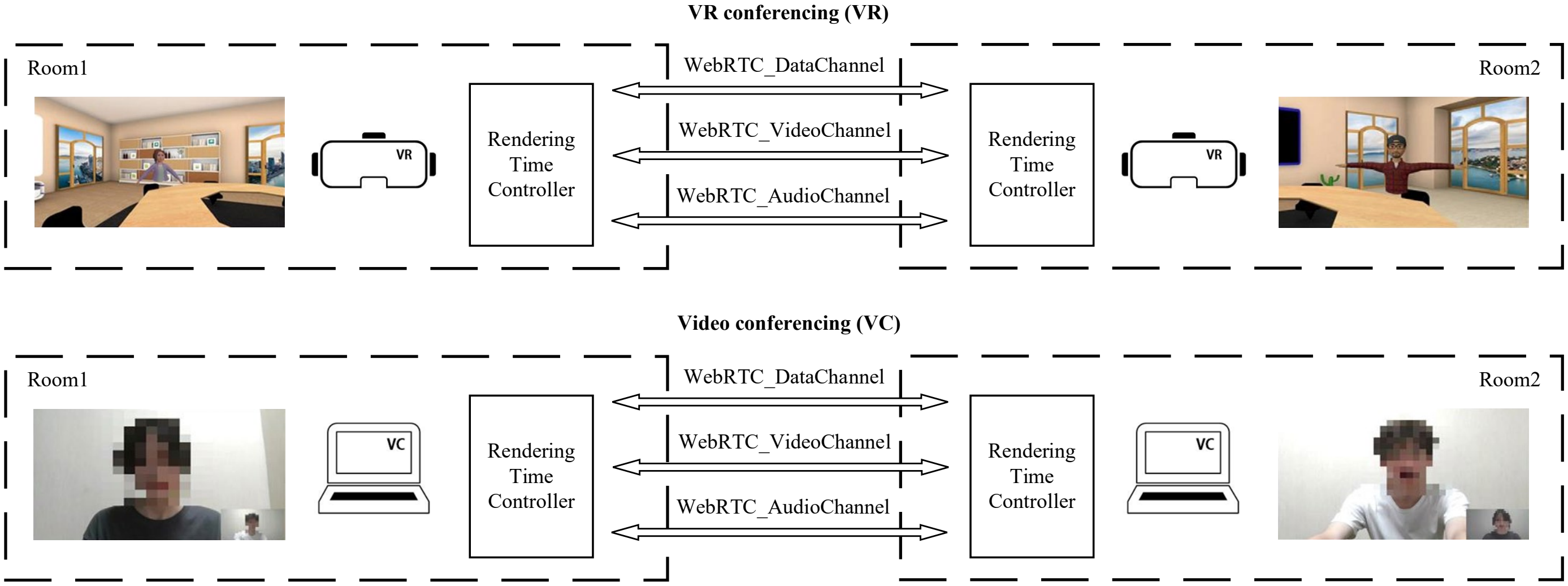

**Fig. 3** Developed applications for VR conferencing and traditional PC-based video conferencing.

*B. Participants*

Each subjective experiment involved 22 participants (7 females and 15 males, ages 23-25, M = 23.64, SD = 0.727), all of whom were university students with normal or corrected-to-normal vision and hearing. The participants were selected because they were relatively young and were more likely to use VR systems. Indeed, all participants reported previous experience with VR headsets as well as traditional PC-based VC systems. According to ITU-T P.920 recommendation [24], participants who are familiar with each other can stimulate more natural and lively conversations. Therefore, each experimental group consisted of friends or acquaintances to encourage active conversational engagement and avoid additional subjective delays due to unfamiliarity or social discomfort.

As this study represents an exploratory investigation of the perceptual-cognitive relationship under network latency, a demographically homogeneous sample was employed to control variability. We expect the identified interaction patterns to be qualitatively applicable across broader user populations, which will be further examined in future studies involving more diverse participant groups.

*C. Experimental procedures*

***Experiment 1*** This experiment aims to study the impact of E2E latency on interaction fluency in VR conferencing and compare it with a traditional VC. The subjective tests included a pre-experiment stage (about 30 min) and a formal experiment stage (about 90 min), and the entire test process lasted about 120 minutes. In the pre-experimental stage, the researchers explained the test procedure and precautions to participants. The participants were then asked to wear VR headsets and enter the shared virtual conferencing environment in two separate rooms, allowing them to become familiar with the virtual environment before the experiments to avoid bias during the formal experiment. Similarly, for traditional VC, participants joined a conference through two laptops to familiarize themselves with the custom application before starting the formal experiment. Participants were asked to use two test platforms for conversation. However, VR conferencing was relatively new to the participants in this experiment. Therefore, they were required to wear HMD and freely explore the shared virtual environment for more than 10 minutes to fully familiarize themselves with the immersive experience and avoid distractions during the main experiment. In addition, participants needed to get used to the VR controllers and gesture tracking function to ensure that they could perform the required tasks during the VR conferencing tests. During the pre-experiment stage, they were also exposed to different levels of latency in both VR and VC conferencing.

In the formal experiment, participants were divided into pairs, for a total of 11 groups. Each group performed two subjective tests: one in an immersive VR conferencing system and the other in a PC-based video conferencing system. Both tests within each group followed the same evaluation criteria. Each test had 7 latency settings: 100ms, 200ms, 300ms, 500ms, 800ms, 1000ms, 1500ms, 2000ms, 2500ms, 3000ms. The order of these latency settings was randomly presented to each pair of participants. To offset potential order effects, the order of the two subjective test modalities also alternated between groups. For instance, if one group began with an immersive VR conferencing test followed by a VC test, the next group would start with a VC test and then proceed to a VR test. Participants had a 20 min break between the two tests.

In each test of the subjective experiment, both participants were subjected to the same experimental parameters, meaning they experienced the same E2E latency. Before the test began, both parties decided who would initiate the conversation. The formal test was divided into three phases: negotiation phase, conversation phase, and rating phase. At the negotiation phase, the participants communicated to confirm whether they were ready to begin the test. Once they reached an agreement, the conversation test began. During the conversation phase, the experiment ensured that all conversations followed a strict "request-response" pattern to simplify the complexity of measuring the intensity of user interaction. Based on this, participants were required to follow a key rule throughout the experiment: they could only speak again after receiving a response from the other party. Additionally, participants could see each other during the conversation to ensure a more natural and immersive interaction. Each conversation test lasted approximately 2.5 minutes, after which a scoring interface appeared on the device's screen. It should be noted that the 2.5-min conversational segments were intended for standardized assessment rather than familiarization, as participants had already completed pre-experimental training and environment acclimation, and each pair consisted of friends or acquaintances with prior social familiarity. At the rating phase, participants were asked to rate the scores of interaction fluency using the Absolute Category Rating (ACR) method, with a labeled 5-point rating scale ("excellent," "good," "fair," "poor," and "bad" fluency corresponding to the scores of 5 to 1) [61]. After that, participants rested for 2.5 mins before the next test. In the subjective test of VR conferencing, participants had to take off their headsets during breaks to prevent motion sickness, which could interfere with the experiment if the headset was worn for too long.

In order to more realistically characterize the impact of E2E latency on conversation, three different types of conversation tasks were designed in this experiment: (1) Counting, (2) Arithmetic, and (3) Free conversation. In the counting task, participants took turns counting sequentially from 1. This task simulated a high-frequency interactive conversation that required no deep thinking, such as rapid-fire Q&A or an intense debate in a high-pressure environment. In the arithmetic task, participants asked each other simple addition questions with sums under 10. After one person answered correctly, the other person continued to ask questions, and they took turns. This task simulated a conversation scenario with a medium interaction frequency, such as the question-and-answer session in online teaching or quizzes. In the free conversation task, participants could choose topics from daily life (such as weather, diet, or travel plans) to discuss. Except for basic instructions, the experimenter did not intervene. This type of natural conversation accounts for the largest proportion in daily communication.

After collecting the subjective scores of 22 participants under different tests, the mean opinion score (MOS) of each test was calculated [61]. The experimental results were screened by calculating the Pearson correlation coefficient (PCC) between each participant's subjective score and the MOS. Participants with a PCC value lower than 0.75 were excluded from the analysis. All participants met this criterion.

***Experiment 2*** This experiment aims to investigate the impact of E2E latency on social presence in VR conferencing and traditional VC. Compared with interaction fluency at the perceptual level, social presence is more complex at the cognitive level, involving the understanding of intention, emotion, and behavior. To assess the social presence, the Networked Minds Social Presence Inventory (NMSPI) with all its items (see Appendix for details) [62] was applied in the experiment. This questionnaire is widely used for measuring social presence, and its reliability and validity have been verified [62]. Additionally, the question items are designed with clear and concise language and a well-structured logical flow to ensure participants can easily comprehend and respond accurately [63]. At the same time, it follows the design principles of cognitive psychology, such as utilizing similarity and consistency in question settings to reduce the cognitive load on subjects [64]. The NMSPI evaluates six subscales of social presence, namely: co-presence (CP), attentional allocation (AA), message understanding (MU), affective understanding (AU), emotional interdependence (EI), and behavioral interdependence (BI). More specifically, CP refers to the user's feeling of not being alone or isolated and their sense of the other's awareness of their presence. AA addresses the degree of attention that users give and receive from their interactants. MU reflects the user's ability to comprehend the messages received from the interactant, as well as their sense of the interactant's understanding of their messages. AU concerns the user's ability to recognize the interactant's emotional and attitudinal states, along with their sense of the interactant's ability to understand their own emotion and attitudes. EI measures the degree to which the user's emotional and attitudinal state affects and is affected by those of the interactant. BI evaluates how the user's behavior affects and is affected by the interactant's behavior [62]. Each subscale has 6 questions. Therefore, each test has a total of 36 questions. All items of the questionnaire adopt a 7-point Likert scale (e.g., Strongly disagree - Strongly agree) [62].

Considering that social presence focuses on assessing the user's cognitive process of being with others and the degree of understanding of other's intentions and emotions in a mediated environment, therefore, this experiment does not consider tasks with clear intentions but a lack of emotions, such as turn counting and simple arithmetic. The subjective experiment followed a similar process to Experiment 1, divided into two stages: the pre-experiment stage and the formal experiment stage, with a total duration of 120 minutes. The pre-experiment lasted around 30 minutes, while the formal experiment took approximately 90 minutes. In the pre-experiment, researchers introduced the test process to the participants and explained the six key concepts in the NMSPI questionnaire. Participants were also asked to have a free conversation under VR and VC conditions, each with different degrees of latency, to experience its impact on their interactions. In the formal experiment, the participants were grouped in pairs. Each pair conducted two subjective tests: one was conducted in an immersive VR conferencing system and the other was conducted in a PC-based VC system. Both tests focused on free conversations. Before the test began, the participants could determine the topic of the conversation by themselves. The rule of the entire experiment was that participants could only respond after the other party had completed their statement. Both tests followed the same latency level, with seven latency settings arranged in random order: 100ms, 500ms, 1000ms, 1500ms, 2000ms, 2500ms, and 3000ms. To counteract potential order effects, the order of the two tests was alternated between sessions. For each individual test, both participants received the same E2E latency settings. Each test lasted approximately 2.5 minutes and was followed by a 2.5-minute rest during which participants completed a questionnaire on the computer. During the rest periods, participants removed the HMD to prevent visual fatigue and cybersickness. In addition, Experiments 1 and 2 were conducted in separate sessions rather than consecutively.

*D. Bias control*

In the experiment, potential confounding factors that could contribute to the observed differences between VC and VR, such as device characteristics, technological familiarity, expectation effects, and possible social desirability bias, were carefully considered in the experimental design. A within-subject study design was adopted, where all participants experienced both VC and VR conditions using fixed hardware configurations. Pre-experimental training and acclimation sessions were provided to normalize participants' familiarity with both platforms and to reduce novelty effects. In addition, questionnaires were completed individually and anonymously, and participants were not informed of any expected outcomes or preferred conditions, thereby mitigating potential experimenter-pleasing biases. While subtle residual effects cannot be completely excluded, these controls minimize their influence and allow the observed differences to be primarily attributed to system- and latency-related factors.

## IV. Experimental Results and Analysis

In this section, we analyze the effects of E2E latency on both interaction fluency and social presence in VR conferencing, and compare with traditional VC based on the subjective experimental results.

*A. Impact of Latency on Interaction Fluency*

The impact of latency on interaction fluency under VC and VR will be first analyzed, respectively. Then a comparison will be conducted to check the difference of the impact between the two conferencing sceneries.

*(1) Interaction Fluency Analysis for VC*

Fig. 4 illustrates the relationship between E2E latency and interaction fluency in VC under different conversation tasks. In general, for the three types of conversation tasks, the interaction fluency scores (in terms of MOS) will all decrease as the E2E latency increases. However, there are obvious differences in the changing trends of interaction fluency with E2E delay between different conversation tasks.

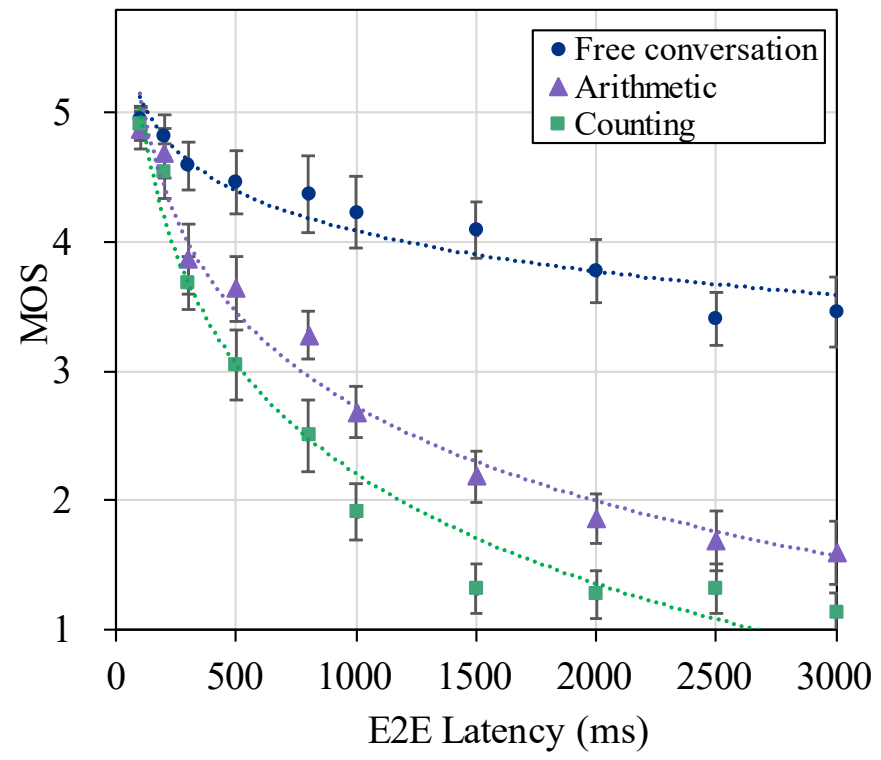

**Fig. 4** Relationship between E2E latency and interaction fluency under different conversation tasks for VC.

In the free conversation task, as E2E latency increases from 100ms to 1,500ms, interaction fluency gradually declines from nearly 5 to around 4, representing a total drop of 1. Even with latency rising to 3,000ms, the interaction fluency remains relatively stable at approximately 3.5. In contrast, during the arithmetic task, the interaction fluency declines more rapidly. When E2E latency reaches 300ms, interaction fluency drops to around 4 or lower. Following this trend, the fluency drops to 3 when latency reaches approximately 800ms and further drops to about 2 when the latency increases to 2,000 ms. Compared to the free conversation task, the arithmetic task shows a steeper decline in interaction fluency, with a total drop of nearly 3 points as latency increases from 100 ms to 2,000 ms. This indicates that interaction fluency is significantly more sensitive to latency during the arithmetic task than during the free conversation task. In the counting task, as the E2E delay increases, the trend of interaction fluency decreases is similar to that of the simple arithmetic task, but the fluency decreases faster. When latency exceeds 100ms, the interaction fluency score in the counting task is always lower than that in the arithmetic task. Based on the decreasing trend of interaction fluency in the counting task, it can be estimated that when the latency increases to 500ms, the interaction fluency will further decrease to about 3. When the latency reaches 1,500ms, the interaction fluency will drop to about 1.3. In the process of increasing the latency from 100ms to 1500ms, the interaction fluency decreases by nearly 3.7 in total.

Among the three tasks, E2E latency has the greatest impact on interaction fluency in the counting task, a moderate impact in the arithmetic task, and the least impact in the free conversation task. When E2E latency is below 100ms, users seem not sensitive to it, resulting in minimal differences in interaction fluency across tasks. However, as latency increases, users become more aware of it, and the variations in latency's impact on interaction fluency across tasks become more obvious. In practice, when users interact through a system, their interaction fluency is influenced by two main factors: (1) individual characteristics, such as personality, prior experience, perception, involvement, motivation, and behavior, and (2) the performance of the system providing the interactive function. The degradation of the interaction fluency caused by latency may be introduced either by the system itself or by the other person. The concept of "attribution of latency" arises when users perceive a noticeable response latency and must determine whether it is caused by system-related latency or by the subjective delay of the other person.

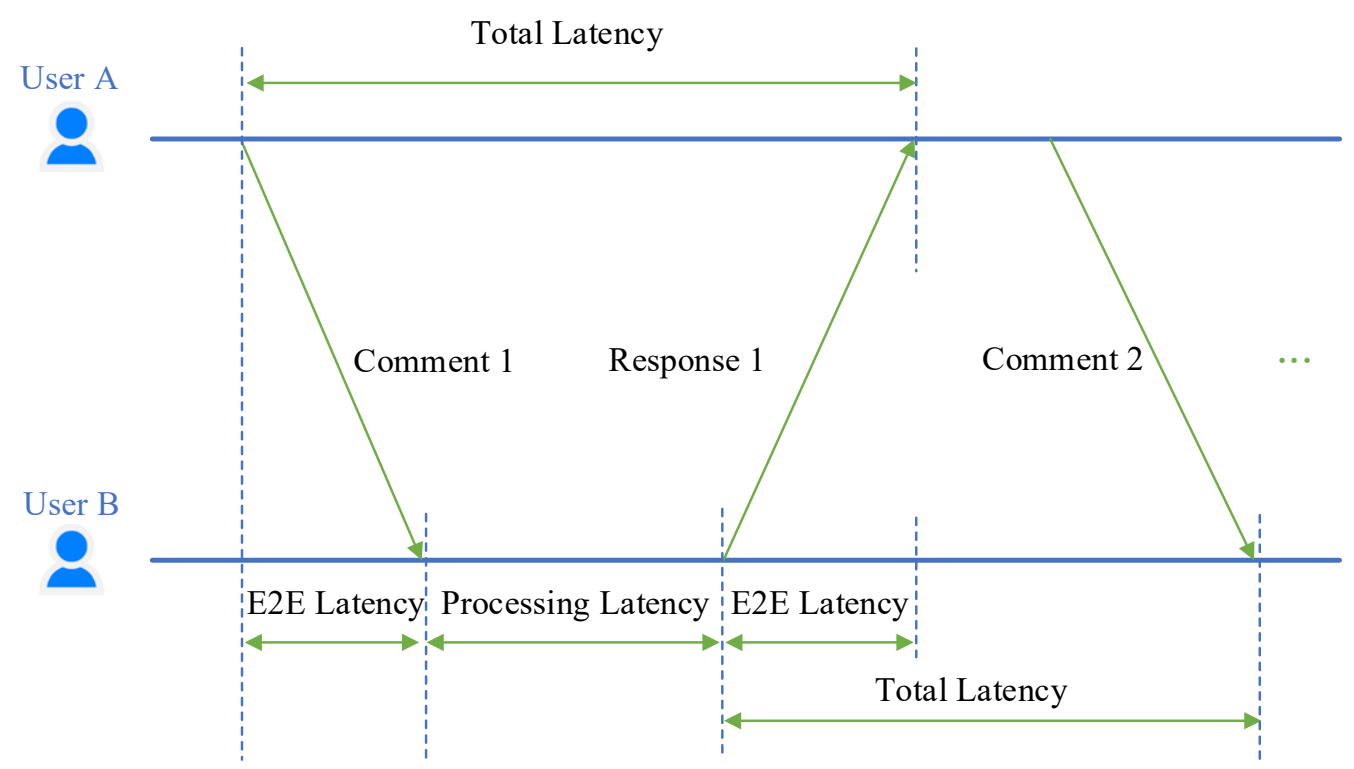

**Fig. 5** Latency during the user conversation.

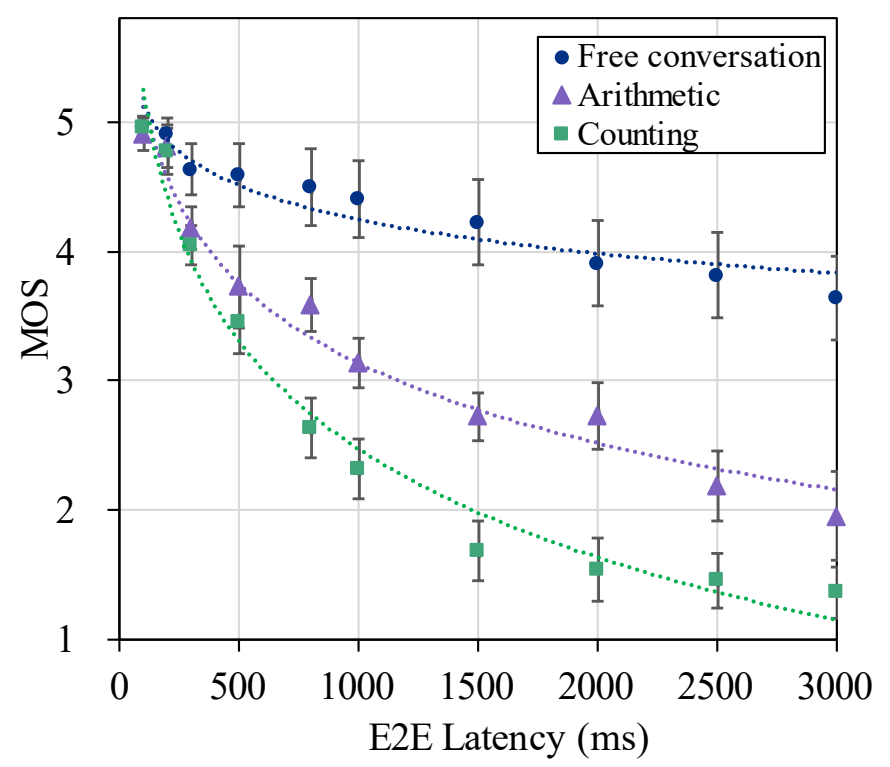

**Fig. 6** Relationship between E2E latency and interaction fluency under different conversation tasks in VR conferencing.

Compared to the free conversation task, the counting and simple arithmetic tasks are easier to understand, requiring less effort to maintain interaction. When users feel that the conversation is delayed, they are more likely to attribute a decline in interaction fluency to the system performance degradation. Drawing from their FTF interaction experience, users are more tolerant of response latency during free conversation and tend to attribute such latency to their party rather than the system. Therefore, in mediated communication systems, factors such as the "attribution of latency" may also influence interaction fluency.

To better clarify this concept, we further analyze the latency experienced within a conversational interaction. Fig. 5 illustrates the end-to-end timing structure during user communication. When User A initiates a message, it reaches User B after an E2E latency. User B then cognitively processes the content and formulates a response, which incurs a processing latency. The response is subsequently transmitted back to User A with an additional E2E latency. Accordingly, the overall latency within a conversational turn consists of three components, namely, the E2E latency from User A to User B, the user processing latency, and the E2E latency from User B to User A. Assuming symmetric transmission delays, the total latency is the sum of the two E2E latencies and the processing latency, which can be expressed as follows:

$$L = 2L_E + L_P \tag{1}$$

where $L$ is the total latency, $L_E$ is the E2E latency, and $L_P$ is the processing latency. The proportion $\delta$ of E2E latency in the total latency is:

$$\delta = \frac{2L_E}{2L_E + L_P} = \frac{1}{1+\mu} \tag{2}$$

where $\mu$ is expressed as:

$$\mu = \frac{L_P}{2L_E} \tag{3}$$

Accordingly, the ratio $\delta$ represents the contribution of E2E latency to the total latency perceived by the user and reflects the relative influence of transmission delay on interaction fluency. A larger $\delta$ indicates a stronger impact of E2E latency and a more pronounced reduction in interaction fluency. The value of $\delta$ depends on users' processing and reaction time, which varies across conversation tasks. For instance, in the counting task, users respond almost immediately without needing to process, resulting in a low $L_P$. In the calculation task, users require some information processing to respond, though not excessively, yielding an $L_P$ higher than that in the counting task. For the free conversation task, the user needs to think about the other party's questions and organize the language to reply, resulting in a longer thinking time and the highest $L_P$ value compared to the other tasks. Correspondingly, when the E2E latency is fixed, $\delta$ is largest in the counting task, moderate in the calculation task, and smallest in the free conversation task. This indicates that E2E latency has a smaller impact in the free conversation task, whereas its effect is more pronounced in the counting task.

*(2) Interaction Fluency Analysis for VR*

Fig. 6 illustrates the relationship between E2E latency and interaction fluency in VR conferencing across different conversation tasks. Similar to video conferencing, interaction fluency in VR decreases as E2E latency increases across all three task types. However, there are also obvious differences in the trend of interaction fluency between different conversation tasks as the E2E latency changes.

In the free conversation task, when the E2E latency increases from 100ms to 2000ms, the interaction fluency gradually decreases from nearly 5 to about 4, resulting in a total decrease of 1. In contrast, from the downward trend of the interaction fluency in the arithmetic task, when the E2E latency increases to 400ms, the interaction fluency will drop to about 4. As the latency increases to 1,200ms, the MOS score will drop to about 3, and when the delay increases to 2,000ms, the score will further drop to about 2.5. Compared with the free conversation task, the interaction fluency in the arithmetic task decreases by nearly 2.5 in total as E2E latency increases from 100ms to 2,000ms. The interaction fluency decreases faster, which also shows that interaction fluency in the arithmetic task is more sensitive to latency.

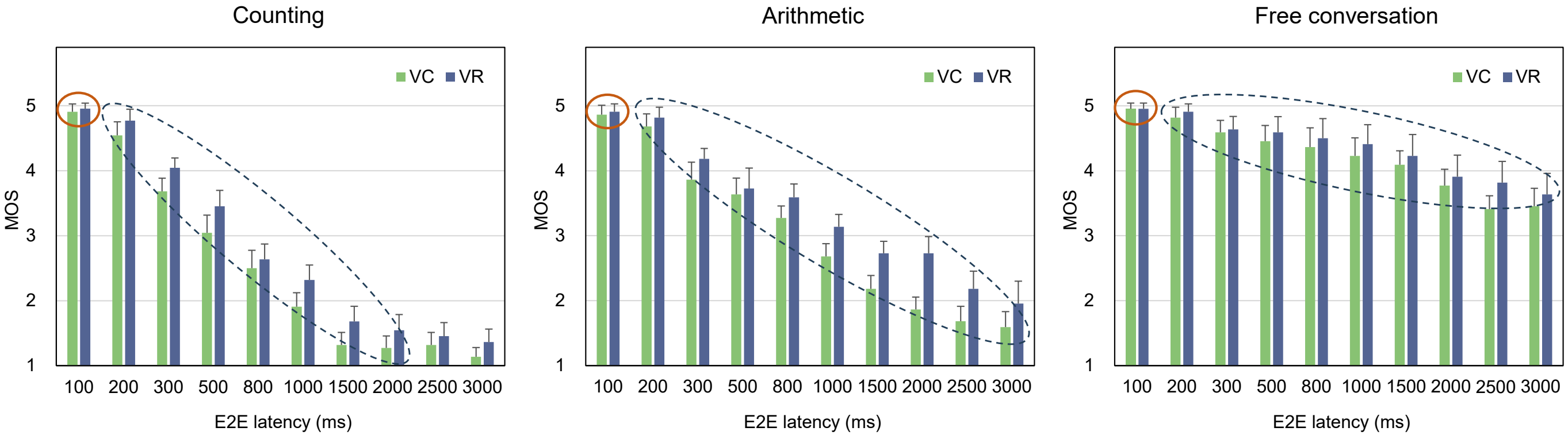

**Fig. 7** Comparison between E2E latency and interaction fluency under different conversation tasks for VC and VR conferencing.

In the counting task, as the E2E delay increases, the trend of interaction fluency decreases is similar to that of the simple arithmetic task, but the fluency decreases faster. When the latency exceeds 200ms, the interaction fluency score in the counting task is always lower than that in the arithmetic task. According to the decreasing trend of the interaction fluency in the counting task, it can be estimated that when the latency is 300ms, the interaction fluency will drop to about 4. When the latency increases to 600ms, the interaction fluency will further drop to about 3. When the latency reaches 2,000ms, the interaction fluency will drop to about 1.5. In the process of increasing the latency from 100ms to 2,000ms, the interaction fluency decreases by nearly 3.5 in total. Among the three tasks, E2E latency has the greatest impact on interaction fluency in the counting task, a moderate impact in the arithmetic task, and the least impact in the free conversation task.

*(3) Interaction Fluency Comparison between VC and VR*

Fig. 7 shows the interaction fluency of different tasks in both VC and VR conferencing scenarios. As the latency increases, the interaction fluency of users in VR and VC conferencing shows a downward trend, but the differences in interaction fluency vary across different conversation tasks and latency levels. Specifically, when the E2E latency is 100ms, as marked by the orange solid box in the figure, the interaction fluency in VR and VC conferencing does not show obvious differences in all tasks (T-test, $p<0.05$), and both remain at a high score (close to 5). This result reveals that in a conversation task, when the network is in an ideal state, the perceived fluency of VR and VC conferencing is basically the same for users. Additionally, users do not seem to be sensitive to a delay of 100ms, as they can successfully complete the conversation task even in the more intensive counting task.
However, when the E2E latency exceeds 200ms, as marked by the blue dashed box in the figure, the latency becomes noticeable and starts to impact user interaction. In this case, the interaction fluency in VR and VC conferencing shows a significant difference for all three conversation tasks (T-test, $p<0.05$), with VR conferencing generally providing a higher interaction fluency than VC. This result indicates that E2E latency affects the interaction fluency of VR and VC conferencing differently, with users exhibiting a higher tolerance for latency in VR conferencing. Moreover, the difference between VC and VR conferencing varies across task scenarios. In higher-intensity tasks, such as counting and arithmetic, latency leads to a more pronounced decline in interaction fluency for VC compared to VR. Consequently, the disparity in interaction fluency between VC and VR is more noticeable in these tasks than in the free conversation task. This result indicates that users in VC are more sensitive to fluency in high-intensity conversations, making them more affected by latency. In contrast, in low-intensity tasks like free conversation, the difference in the impact of latency on VC and VR conferencing is relatively small.

This difference between VR and VC may be attributed to the following reason: compared with traditional VC, where the user's FOV displays a real person's image, the VR conferencing here uses an animated digital avatar. Interaction with a digital avatar (controlled by a real person) is still different from interaction with a real person. Since digital representation requires users to invest more cognitive resources, thereby reducing their perceptual sensitivity (this effect is more obvious in counting and simple arithmetic tasks with low cognitive demands), they are more tolerant of the decrease in interaction fluency caused by E2E latency.

*B. Impact of Latency on Social Presence*

Before analyzing the impact of E2E latency on social presence, the internal consistency of each inventory subscale in the subjective experimental results is checked using Cronbach's alpha coefficient [65]. The findings indicate that the rating scores in each subscale exhibit excellent internal consistency, with Cronbach's $\alpha$ values ranging from 0.912 to 0.968. Next, the Shapiro-Wilk test [66] is applied to determine whether participant scores follow a normal distribution. Results reveal that scores for CP, AA, MU, EI, and BI are normally distributed, whereas scores for AU are not. Based on these distributions, the impact of latency on social presence in

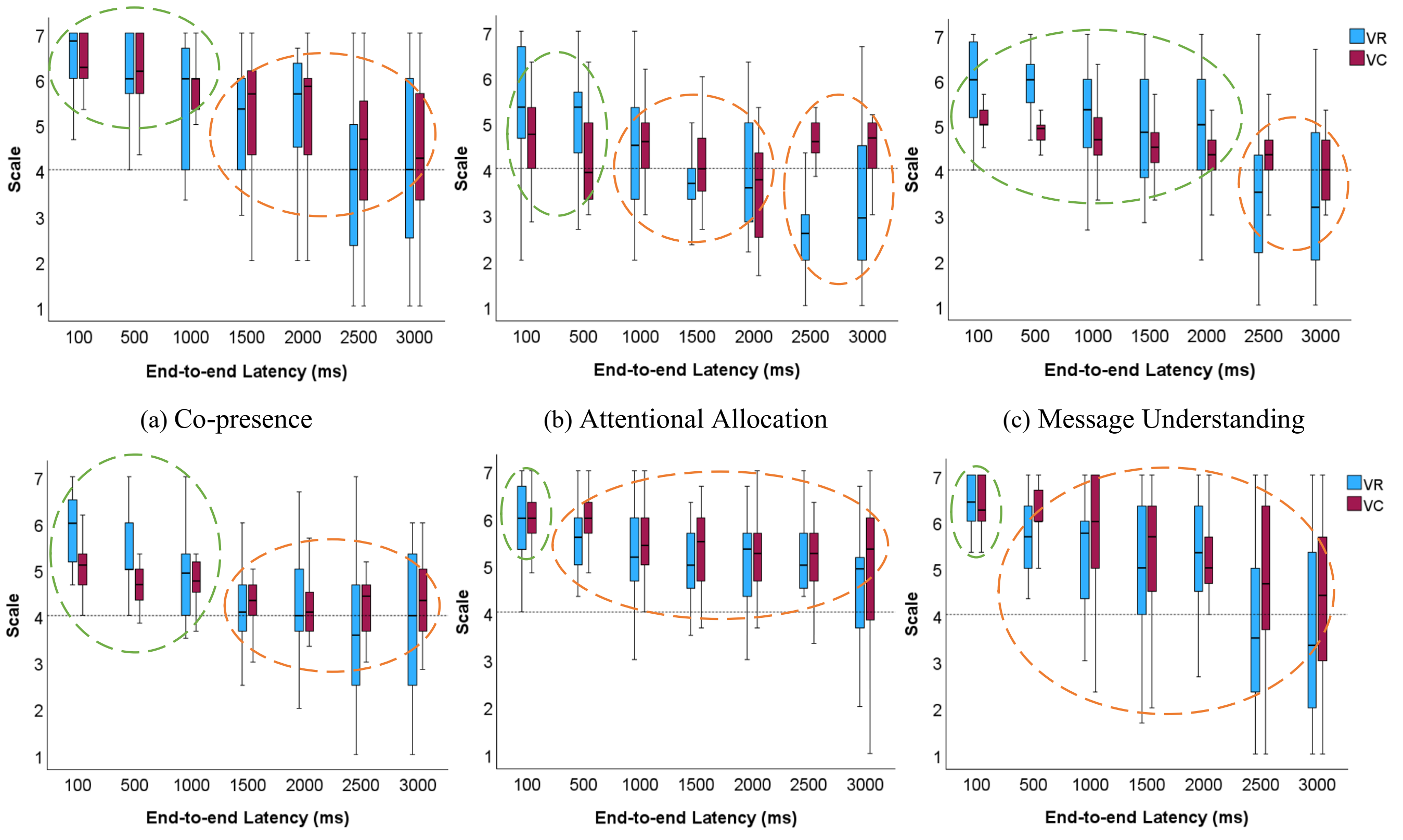

(a) Co-presence (b) Attentional Allocation (c) Message Understanding

(d) Affective Understanding (e) Emotional Interdependence (f) Behavioral Interdependence

**Fig. 8**. Evaluation of the social presence using six subscales and a Likert scale rating (1: Strongly disagree - 7: Strongly agree). From left to right and top to bottom: (a) co-presence (CP), (b) attentional allocation (AA), (c) message understanding (MU), (d) affective understanding (AU), (e) emotional interdependence (EI), and (f) behavioral interdependence (BI). The green dotted line indicates that VR is higher than VC, and the orange dotted line indicates that VR is lower than VC.

VR and VC settings is then compared. Specifically, for score differences that meet the normality assumption, a paired sample T-test [67] is carried out to further check the latency impact differences between VR and VC conferencing. For score differences that do not meet this assumption, a Wilcoxon signed-rank test [68] is conducted instead of the T-test. All significance tests are conducted with a two-tailed α-level of 0.05. Each inventory subscale is analyzed in detail as follows:

Co-presence (CP): As shown in Fig. 8a, the central tendencies of the CP scores suggest that users experienced a decrease in CP as E2E latency increased in both VR and VC conferencing. Paired-sample T-test results ($T = -1.057$, $p>0.05$) indicate no significant difference in the degradation of CP due to E2E latency between the two conference sceneries. This means that the impact of E2E latency on CP is essentially consistent across both VR and VC conferencing scenarios.

Attentional Allocation (AA): In VR conferencing, as shown in Fig. 8b, the central tendency of AA scores declines as the E2E latency increases, while no such trend is observed in VC. T-test results indicate that there is a significant difference between VR and VC in the decrease in AA due to E2E latency ($T=-2.055$, $p<0.05$). Specifically, when E2E latency is below 500ms, AA in VR conferencing is higher than that in VC. However, as the latency continues to increase, AA in VR conferencing gradually drops lower than that in VC, and this difference becomes more obvious when the latency exceeds 2,500ms. These results indicate that although VR can provide users with higher AA at low latency, AA is more sensitive to increases in latency. Therefore, strict latency control is required to maintain AA in VR conferencing.

Message Understanding (MU): As shown in Fig. 8c, E2E latency has a negative impact on the user's MU in both VR and VC conferencing. According to the T-test results, there is also a significant difference in the MU degradation due to the increase in latency between VR and VC ($T=2.812$, $p<0.05$). Fig. 8c further illustrates that as E2E delay increases, MU in VR decreases faster than in VC, indicating that MU in VR is more sensitive to latency. In addition, when E2E latency is below 2,000 ms, the MU score in VR is higher than that in VC However, once the latency exceeds 2,500ms, the MU in VR drops sharply and eventually becomes lower than that in VC. Similar to AA, the results for MU indicate that while VR conferencing can initially support better MU than VC, it also requires strict latency control to maintain this advantage.

Affective Understanding (AU): According to the Wilcoxon signed rank test results ($Z=-0.567$, $p=0.571$), there is no significant difference between the relationships of the AU and E2E latency for VR and VC. However, as shown in Fig. 8d, when the delay increases from 100ms to 1500ms, the AU gradually decreases, but after exceeding 1500ms, the AU no longer continues to decrease. When the E2E delay is less than 500ms, the AU of VR is higher than that of VC. These results show that VR conferencing can provide users with better AU than VC if latency is kept at a low level.

Emotional Interdependence (EI): As illustrated in Fig. 8e, EI gradually decreases as the latency increases from 100ms to 1,000ms in both VR and VC, but does not continue to decrease beyond 1,000ms. The results of the T-test reveal that the decrease in EI due to E2E latency is significantly different between VR and VC (T=-2.032, $p<0.05$). It can be found that EI is higher in VC than in VR for most of the considered latency values. These findings suggest that VR does not provide users with better EI than VC, possibly because avatars in current VR systems [14] (including our system) still do not accurately reflect the user's true appearance and expressions, potentially affecting EI.

Behavioral Interdependence (BI): As shown in Fig. 8f, BI gradually decreases with increasing E2E latency in both VR and VC conferencing scenarios. T-test results indicate a significant difference in the BI decrease caused by latency between VR and VC (T=-3.036, $p<0.05$). It can be found that in most of the considered latency values, the BI value in VC is generally higher than that in VR, which indicates that VR does not provide users with better BI than VC. Similar to EI, this may be because the avatars used in current VR conferencing systems [14] do not accurately reflect the user behaviors, thereby affecting BI.

In general, low latency (less than 100ms) has less impact on the cognitive dimension of social presence, allowing VR to achieve the same or even higher effect as VC through its novel interaction and presentation style. However, as latency increases, its impact on the sub-dimensions of social presence becomes more pronounced in VR than in VC, potentially degrading the VR experience. Overall, to maintain a good cognitive level in VR conferencing (with sub-dimension scores remaining above 5 and exceeding those of VC), latency should be kept below 1 s. It should be noted that a relatively large variance was observed in both VC and VR conditions, reflecting substantial inter-individual differences in sensitivity to mediated social interaction and network latency. This variability becomes more pronounced under high-latency settings, further amplifying individual response differences. Such dispersion may be attributed to heterogeneity in cognitive processing styles, conversational habits, and tolerance to interaction disruptions. Importantly, although dispersion increases across conditions, the conclusions drawn above are not based solely on average trends. All major findings are supported by statistical analyses, including reliability assessment, normality testing, and paired-sample significance tests, which explicitly account for participant-level variability. The statistically significant effects observed across multiple sub-dimensions demonstrate that the identified latency-related trends remain robust despite elevated variance, while increased dispersion under extreme conditions highlights the need for future work to investigate individual response patterns and personalized latency tolerance.

TABLE I

COMPARISON PARAMETERS BETWEEN INTERACTION FLUENCY AND EACH SUB-DIMENSION OF SOCIAL PRESENCE

| VR Conferencing | | | | | | |
|---|---|---|---|---|---|---|
| | **CP** | **AA** | **MU** | **AU** | **EI** | **BI** |
| **Slope** | -0.154 | -0.162 | -0.165 | -0.120 | -0.064 | -0.174 |
| **SRCC** | 0.919 | 0.929 | 0.929 | 0.937 | 0.865 | 0.929 |
| **RMSE** | 0.012 | 0.096 | 0.033 | 0.050 | 0.014 | 0.029 |
| VC Conferencing | | | | | | |
| | **CP** | **AA** | **MU** | **AU** | **EI** | **BI** |
| **Slope** | -0.107 | 0.005 | -0.058 | -0.046 | -0.041 | -0.106 |
| **SRCC** | 0.811 | 0.036 | 0.937 | 0.775 | 0.844 | 0.955 |
| **RMSE** | 0.079 | 0.255 | 0.182 | 0.191 | 0.094 | 0.051 |

### *C. Relationship Analysis between Interaction Fluency and Social Presence Under Various Latencies*

After analyzing the individual impacts of latency on interaction fluency and social presence, we further examine the relationship between interaction fluency from the perspective of quality perception and social presence from the perspective of social cognition under varying latency conditions. Considering the different value ranges of interaction fluency and social presence scores, each score was normalized before analysis, and a linear regression was performed on this basis to determine the slope of each score as the latency increased. In addition, the Spearman rank correlation coefficients (SRCC) and root mean square error (RMSE) between interaction fluency and each sub-dimension of social presence were calculated for comparative analysis. The details are shown in Table 1.

For VR conferencing, both interaction fluency and the sub-dimensions of social presence exhibit a declining trend as latency increases (all slopes are negative). Among these, AU demonstrates a decay rate (slope= -0.120) most closely aligned with the decay rate of interaction fluency (slope = -0.110). The correlation between AU and interaction fluency is the highest as well (correlation coefficient = 0.937, $p<0.05$). In contrast, the slopes for CP, AA, MU, and BI are smaller than that of interaction fluency, indicating these dimensions decay faster than interaction fluency (correlation coefficients> 0.9, $p < 0.05$). This suggests that when a decline in fluency at the perceptual level, it may have a more pronounced impact on CP, AA, MU, and BI at the cognitive level. Interestingly, while EI also declines as interaction fluency decreases, its decay rate is slower than that of interaction fluency, and its correlation with interaction fluency is lower compared to other sub-dimensions. A possible explanation is that EI is built on long-term interactions, and short-term latency issues do not significantly impact the emotional connections between users. These findings underscore the importance of these dimensions in shaping the overall user experience in VR conferencing.

For VC conferencing, the relationship between interaction fluency and the sub-dimensions of social presence under increasing latency differs somewhat from that in VR conferencing. In detail, BI (slope = -0.106) and CP (slope = -0.107) exhibit decay rates most similar to that of interaction fluency (slope = -0.129). However, their decay rates are still slightly lower than that of interaction fluency. This suggests that the perception of interaction fluency significantly influences the user's cognitive processes of CP and BI in VC scenarios. This effect may be due to the lack of coordination of actions caused by latency, which weakens the sense of the other partner being "present in real time". In contrast, the slopes of MU, AU, and EI are significantly smaller than the slope of interaction fluency. This implies that while declines in interaction fluency do affect these dimensions, the impact is limited. The slower decay of EI is consistent with the findings in VR scenarios, where EI seems to rely on long-term interactions. For MU and AU, although latency affects real-time interaction, it does not entirely undermine verbal and non-verbal cues, which can be supplemented by delayed cues in VC settings. Notably, AA does not show a significant decline as interaction fluency decreases (slope=0.005>0, correlation coefficient=0.036, p>0.05). This suggests that in traditional VC, the user's attention remains largely focused on the conversation tasks even in high-latency conditions. This contrasts with VR conferencing, where high latency often distracts users, leading them to focus on other surrounding elements.

## V. Conclusion and Further Discussion

This study examined the impact of latency on interaction fluency and social presence in VR conferencing and compared it with traditional VC. The findings reveal that the effect of latency on interaction fluency varies across conversational tasks. In less cognitively demanding tasks, users are more prone to attribute latency to system performance issues, whereas in tasks imposing greater cognitive load on the other party, they tend to attribute latency to their conversation partner. Moreover, as latency increases, users appear less sensitive to declines in interaction fluency in VR conferencing. This may relate to interaction with digital avatars, where users allocate more cognitive resources, resulting in higher tolerance for latency-induced fluency degradation. Regarding social presence, VR conferencing shows superior cognitive capacity at low latency but falls behind VC at higher latency levels. This contrasts with the consistently superior interaction fluency of VR conferencing across different latency conditions. These findings highlight the dual nature of VR conferencing: while VR mitigates the perceptual impact of latency on interaction fluency, it may impose a greater cognitive burden, especially as latency increases, leading to a more significant decline of social presence compared to VC.

The current study is primarily exploratory, aiming to investigate the differential impact of latency on perception and cognition. Future research will extend the evaluation to more diverse populations to validate generalizability and examine how increased VR exposure and more expressive digital avatars (e.g., with realistic facial expressions and appearances) affect the impact of latency on perception and cognition.

## Appendix

Networked Minds Social Presence Inventory (NMSPI) [55]

<table>
<tr><th>Factor Items</th><th>ID</th><th>Questionnaire Items</th></tr>
<tr><td rowspan="6">Co-presence</td><td>A1</td><td>I noticed my partner</td></tr>
<tr><td>A2</td><td>My partner noticed me</td></tr>
<tr><td>A3</td><td>My partner's presence was obvious to me</td></tr>
<tr><td>A4</td><td>My presence was obvious to my partner</td></tr>
<tr><td>A5</td><td>My partner caught my attention</td></tr>
<tr><td>A6</td><td>I caught my partner's attention</td></tr>
<tr><td rowspan="6">Attentional Allocation</td><td>B1</td><td>I was easily distracted from my partner when other things were going on</td></tr>
<tr><td>B2</td><td>My partner was easily distracted from me when other things were going on</td></tr>
<tr><td>B3</td><td>I remained focused on my partner throughout our interaction</td></tr>
<tr><td>B4</td><td>My partner remained focused on me throughout our interaction</td></tr>
<tr><td>B5</td><td>My partner did not receive my full attention</td></tr>
<tr><td>B6</td><td>I did not receive my partner's full attention</td></tr>
<tr><td rowspan="6">Message Understanding</td><td>C1</td><td>My thoughts were clear to my partner</td></tr>
<tr><td>C2</td><td>My partners thoughts were clear to me</td></tr>
<tr><td>C3</td><td>It was easy to understand my partner</td></tr>
<tr><td>C4</td><td>My partner found it easy to understand me</td></tr>
<tr><td>C5</td><td>Understanding my partner was difficult</td></tr>
<tr><td>C6</td><td>My partner had difficulty understanding me</td></tr>
<tr><td rowspan="6">Affective Understanding</td><td>D1</td><td>I could tell how my partner felt</td></tr>
<tr><td>D2</td><td>My partner could tell how I felt</td></tr>
<tr><td>D3</td><td>My partner's emotions were not clear to me</td></tr>
<tr><td>D4</td><td>My emotions were not clear to my partner</td></tr>
<tr><td>D5</td><td>I could describe my partner's feelings accurately</td></tr>
<tr><td>D6</td><td>My partner could describe my feelings accurately</td></tr>
<tr><td rowspan="6">Emotional Interdependence</td><td>E1</td><td>I was sometimes influenced by my partner's mood</td></tr>
<tr><td>E2</td><td>My partner was sometimes influenced by my mood</td></tr>
<tr><td>E3</td><td>My partner's feelings influenced the mood of our interaction</td></tr>
<tr><td>E4</td><td>My feelings influenced the mood of our interaction</td></tr>
<tr><td>E5</td><td>My partner's attitude influenced how I felt</td></tr>
<tr><td>E6</td><td>My attitude influenced how my partner felt</td></tr>
<tr><td rowspan="6">Behavioral Interdependence</td><td>F1</td><td>My behavior was often in direct response to my partner's behavior</td></tr>
<tr><td>F2</td><td>The behavior of my partner was often in direct response to my behavior</td></tr>
<tr><td>F3</td><td>I reciprocated my partner's actions</td></tr>
<tr><td>F4</td><td>My partner reciprocated my actions</td></tr>
<tr><td>F5</td><td>My partner's behavior was closely tied to my behavior</td></tr>
<tr><td>F6</td><td>My behavior was closely tied to my partner's behavior</td></tr>
</table>

## References

[1] S. N. B. Gunkel, R. Hindriks, K. M. E. Assal, et al, "VRComm: an end-to-end web system for real-time photorealistic social VR communication," in *Proc. 12th ACM MMSys*. 2021, pp.65-79.

[2] J. Lou, Y. Wang, C. Nduka, et al., "Realistic facial expression reconstruction for VR HMD users," *IEEE Trans. Multimedia*, vol. 22, no. 3, pp. 730-743, 2019.

[3] X. Wang, W. Zhang, C. Sandor, et al., "Real-and-present: Investigating the use of life-size 2D video avatars in HMD-based AR teleconferencing," *IEEE Trans. Vis. Comput. Graph.*, vol. 31, no. 9, pp. 5626-5641, 2024.

[4] F. L. Fan, H. Li, and M. Shi, "Redirected walking for exploring immersive virtual spaces with HMD: A comprehensive review and recent advances," *IEEE Trans. Vis. Comput. Graph.*, vol. 29, no. 10, pp. 4104-4123, 2022.

[5] S. Y. Chen, Y. K. Lai, S. Xia, et al., "3D face reconstruction and gaze tracking in the HMD for virtual interaction," *IEEE Trans. Multimedia*, vol. 25, pp. 3166-3179, 2022.

[6] A. F. Di Natale, C. Repetto, D. Villani, "Desktop-based virtual reality social platforms versus video conferencing platforms for online synchronous learning in higher education: An experimental study to evaluate students' learning gains and user experience," *J. Comput. Assist. Learn.*, vol. 40, no. 6, pp. 3454-3473, 2024.

[7] C. Keighrey, R. Flynn, S. Murray, et al., "A physiology-based QoE comparison of interactive augmented reality, virtual reality and tablet-based applications," *IEEE Trans. Multimedia*, vol. 23, pp. 333-341, 2020.

[8] X. Wei, X. Jin, and M. Fan. "Communication in immersive social virtual reality: A systematic review of 10 years' studies," in *Proc. Chinese CHI.* 2022, pp. 27-37.

[9] M. Chessa and F. Solari, "The sense of being there during online classes: Analysis of usability and presence in web-conferencing systems and virtual reality social platforms," *Behav. Inf. Technol.*, vol. 40, no. 12, pp. 1237-1249, 2021.

[10] S. Vlahovic, M. Suznjevic, L. Skorin-Kapov. "A survey of challenges and methods for Quality of Experience assessment of interactive VR applications," *J. Multimodal User Interfaces*, vol. 16, no. 3, pp. 257-291, 2022.

[11] S. Egger, P. Reichl, K. Schoenenberg. "Quality of experience and interactivity," *Quality of Experience: Advanced Concepts, Applications and Methods*. Cham: Springer International Publishing, 2014: 149-161.

[12] C. S. Oh, J. N. Bailenson, G. F. Welch. "A systematic review of social presence: Definition, antecedents, and implications," *Frontiers in Robotics and AI*, vol. 5, pp. 1-35, 2018.

[13] D. Roberts, T. Duckworth, C. Moore, R. Wolff, and J. O'Hare, "Comparing the end to end latency of an immersive collaborative environment and a video conference," in *Proc. 13th IEEE/ACM DS-RT.* IEEE, 2009, pp. 89-94.

[14] R. Cheng, N. Wu, M. Varvello, S. Chen, and B. Han, "Are we ready for metaverse? a measurement study of social virtual reality platforms," in *Proc. 22nd ACM IMC*, 2022, pp. 504-518

[15] M. Mody, P. Swami, P. Shastry. "Ultra-low latency video codec for video conferencing," in *Proc. CONECCT*. IEEE, 2014: 1-5.

[16] M. H. Hajiesmaili, L. T. Mak, Z. Wang, et al. "Cost-effective low-delay design for multiparty cloud video conferencing," *IEEE Trans. Multimedia*, vol. 19, no. 12, pp. 2760-2774, 2017.

[17] Y. Xiao, S. Chen, A. C. Zhou, et al. "Low-Latency Video Conferencing via Optimized Packet Routing and Reordering," in *Proc. 32nd IEEE/ACM IWQoS*. IEEE, 2024: 1-10.

[18] J. Xu, B. W. Wah. "Exploiting just-noticeable difference of delays for improving quality of experience in video conferencing," in *Proc. 4th ACM MMSys*. 2013: 238-248.

[19] J. Skowronek, A. Raake, G. H. Berndtsson, et al. "Quality of experience in telemeetings and videoconferencing: a comprehensive survey," *IEEE Access*, vol. 10, pp. 63885-63931, 2022.

[20] M. S. Anwar, J. Wang, W. Khan, et al., "Subjective QoE of 360-degree virtual reality videos and machine learning predictions," IEEE Access, vol. 8, pp. 148084-148099, 2020.

[21] C. Cortés, P. Pérez, and N. García, "Understanding latency and QoE in social XR," *IEEE Consumer Electronics Magazine*, vol. 13, no. 3, pp. 61-72, May 2024.

[22] C. Cortés, I. Viola, J. Gutiérrez, J. Jansen, S. Subramanyam, E. Alexiou, P. Pérez, N. García, and P. César, "Delay threshold for social interaction in volumetric extended reality communication," *ACM Transactions on Multimedia Computing, Communications, and Applications*, vol. 20, no. 7, pp. 1-22, July 2024.

[23] K. Brunnström, E. Dima, T. Qureshi, M. Johanson, M. Andersson, and M. Sjöström, "Latency impact on quality of experience in a virtual reality simulator for remote control of machines," *Signal Processing: Image Communication*, vol. 89, no. 116005, 2020.

[24] International Telecommunication Union (ITU-T), "Interactive test methods for audiovisual communications," Recommendation ITU-T P.920, 2000.

[25] International Telecommunication Union (ITU-T), "Subjective quality evaluation of audio and audiovisual multiparty telemeetings," Recommendation ITU-T P.1301, 2017.

[26] International Telecommunication Union (ITU-T), "Effect of delays on telemeeting quality," Recommendation ITU-T P.1305, 2016.

[27] International Telecommunication Union (ITU-T), "Subjective test methodologies for 360º video on head-mounted displays," Recommendation ITU-T P.919, 2020.

[28] S. Wang, J. Song, A. Trioux, et al. "Effect of latency on social presence in traditional video conference and VR conference: a comparative study," in *Proc. VCIP*. IEEE, 2023: 1-5.

[29] K. Hornbæk, A. Oulasvirta. "What is interaction?" in *Proc. 2017 CHI.* 2017, pp. 5040-5052.

[30] J. Stromer-Galley, "Interactivity-as-product and interactivity-as-process," *The Information Society*, vol. 20, no. 5, pp. 391-394, 2004.

[31] S. Van Damme, J. Sameri, S. Schwarzmann, et al. "Impact of Latency on QoE, Performance, and Collaboration in Interactive Multi-User Virtual Reality," *Applied Sciences*, vol. 14, pp. 6, pp. 1-26, 2024.

[32] S. Vlahovic, M. Suznjevic, L. Skorin-Kapov. "The impact of network latency on gaming QoE for an FPS VR game," in *Proc. 11th QoMEX*. IEEE, 2019, pp.1-3.

[33] T. Kojic, S. Schmidt, S. Möller, et al. "Influence of network delay in virtual reality multiplayer exergames: Who is actually delayed?" in *Proc. 11th QoMEX*. IEEE, 2019: 1-3.

[34] J. Short, E. Williams, B. Christie. *The social psychology of telecommunications*. Wiley, 1976.

[35] R. Barkhi, V. S. Jacob, H. Pirkul. "An experimental analysis of face to face versus computer mediated communication channels," *Group decision and negotiation*, vol. 8, pp. 325-347, 1999.

[36] J. Weidlich. "Presence at a distance: Empirical investigations toward understanding, modeling, and enhancing social presence in online distance learning environments," FernUniversität Hagen, 2021.

[37] F. Biocca, C. Harms, J. K. Burgoon. "Toward a more robust theory and measure of social presence: Review and suggested criteria," *Presence: Teleoperators & virtual environments*, vol. 12, pp. 5, pp. 456-480, 2003.

[38] S. T. Bulu. "Place presence, social presence, co-presence, and satisfaction in virtual worlds," *Computers & Education*, vol. 58, no.1, pp. 154-161, 2012.

[39] J. Cortese, M. Seo. "The role of social presence in opinion expression during FtF and CMC discussions," *Commun. Res. Rep.*, vol. 29, no. 1, pp. 44-53, 2012.

[40] D. Kim, M. G. Frank, S. T. Kim. "Emotional display behavior in different forms of Computer Mediated Communication," *Comput. Hum. Behav.*, vol. 30, pp. 222-229, 2014.

[41] G. Bente, S. Rüggenberg, N. C. Krämer, et al., "Avatar-mediated networking: Increasing social presence and interpersonal trust in net-based collaborations," *Hum. Commun. Res.*, vol. 34, no. 2, pp. 287-318, 2008.

[42] H. Harris, J. N. Bailenson, A. Nielsen, et al., "The evolution of social behavior over time in Second Life," *Presence: Teleoperators and Virtual Environments*, vol. 18, no. 6, pp. 434-448, 2009.

[43] A. M. Von der Pütten, N. C. Krämer, J. Gratch, et al., "It doesn't matter what you are! Explaining social effects of agents and avatars," *Comput. Hum. Behav.*, vol. 26, no. 6, pp. 1641-1650, 2010.

[44] D. Ahn, Y. Seo, M. Kim, et al., "The effects of actual human size display and stereoscopic presentation on users' sense of being together with and of psychological immersion in a virtual character," *Cyberpsychol. Behav. Soc. Netw.*, vol. 17, no. 7, pp. 483-487, 2014.

[45] S. A. Jin, "The virtual malleable self and the virtual identity discrepancy model: Investigative frameworks for virtual possible selves and others in avatar-based identity construction and social interaction," *Comput. Hum. Behav.*, vol. 28, no. 6, pp. 2160-2168, 2012.

[46] J. Appel, A. von der Pütten, N. C. Krämer, et al., "Does humanity matter? Analyzing the importance of social cues and perceived agency of a computer system for the emergence of social reactions during human-

computer interaction," *Adv. Hum. Comput. Interact.*, vol. 2012, no. 1, p. 324694, 2012.
[47] R. D. Johnson, "Gender differences in e-learning: Communication, social presence, and learning outcomes," *J. Organ. End User Comput.*, vol. 23, no. 1, pp. 79-94, 2011.
[48] Y. H. Cho, S. Y. Yim, and S. Paik, "Physical and social presence in 3D virtual role-play for pre-service teachers," *Internet High. Educ.*, vol. 25, pp. 70-77, 2015.
[49] J. Hauber, H. Regenbrecht, A. Cockburn, and M. Billinghurst, "The impact of collaborative style on the perception of 2D and 3D videoconferencing interfaces," *Open Softw. Eng. J.*, vol. 6, no. 1, pp. 1-20, 2012.
[50] S. Nikolic, M. J. Lee, and P. J. Vial, "2D versus 3D collaborative online spaces for student team meetings: comparing a web conferencing environment and a video-augmented virtual world," in *Proc. 26th Annual Conference of the Australasian Association for Engineering Education*, 2015, pp.1-11.
[51] D. Anton, G. Kurillo, and R. Bajcsy, "User experience and interaction performance in 2D/3D telecollaboration," *Future Gener. Comput. Syst.*, vol. 82, pp. 77-88, 2018.
[52] G. Schwartz, S.-E. Wei, T.-L. Wang, S. Lombardi, T. Simon, J. Saragih, and Y. Sheikh, "The eyes have it: An integrated eye and face model for photorealistic facial animation," *ACM Trans. Graph.*, vol. 39, no. 4, pp. 1-15, 2020.
[53] S. N. Gunkel, S. Dijkstra-Soudarissanane, H. M. Stokking, and O. A. Niamut, "From 2D to 3D video conferencing: Modular rgb-d capture and reconstruction for interactive natural user representations in immersive extended reality (XR) communication," *Front. Signal Process.*, vol. 3: 1139897, pp.1-19, 2023.
[54] Unity real-time development platform, URL https://unity.com/. [Online]. Available: https://unity.com/
[55] Meta horizon workrooms, [Online]. Available: https://www.oculus.com/workrooms/.
[56] Meta quest developer center, [Online]. Available: https://developers.facebook.com/products/oculus/.
[57] Y. Yoon, H. Kim, G. A. Lee, et al., "The effect of avatar appearance on social presence in an augmented reality remote collaboration," in *Proc. 2019 IEEE VR*, IEEE, 2019, pp. 547-556.
[58] R. Deshmukh, N. Nand, A. Pawar, et al., "Video conferencing using WebRTC," in *Proc. 2023 ICSCDS*, IEEE, 2023, pp. 857-864.
[59] B. Garcia, L. Lopez-Fernandez, F. Gortazar, et al., "Analysis of video quality and end-to-end latency in WebRTC," in *Proc. GC Wkshps*, IEEE, 2016, pp. 1-6.
[60] J. Tam, E. Carter, S. Kiesler, et al., "Video increases the perception of naturalness during remote interactions with latency," in *Proc. CHI'12 Extended Abstracts on Human Factors in Computing Systems*, 2012, pp. 2045-2050.
[61] Q. Huynh-Thu, M.-N. Garcia, F. Speranza, P. Corriveau, and A. Raake, "Study of rating scales for subjective quality assessment of high-definition video," *IEEE Trans. Broadcast.*, vol. 57, no. 1, pp. 1-14, 2011.
[62] C. Harms and F. Biocca, "Internal consistency and reliability of the networked minds measure of social presence," in *Proc. Seventh Annual International Workshop: Presence*, Valencia: Universidad Politecnica de Valencia, 2004, pp. 1-7.
[63] D. A. Dillman, J. D. Smyth, and L. M. Christian, *Internet, Phone, Mail, and Mixed-Mode Surveys: The Tailored Design Method*, Indianapolis, Indiana, 2014.
[64] F. J. Fowler Jr, *Survey Research Methods*, Sage Publications, 2013.
[65] D. L. Streiner, "Starting at the beginning: an introduction to coefficient alpha and internal consistency," *J. Pers. Assess.*, vol. 80, no. 1, pp. 99-103, 2003.
[66] S. S. Shapiro and M. B. Wilk, "An analysis of variance test for normality (complete samples)," *Biometrika*, vol. 52, no. 3/4, pp. 591–611, 1965.
[67] H. Hsu and P. A. Lachenbruch, "Paired T-test," Wiley StatsRef: statistics reference online, 2014
[68] R. F. Woolson, "Wilcoxon signed-rank test," *Wiley encyclopedia of clinical trials*, 2007, pp. 1-3.

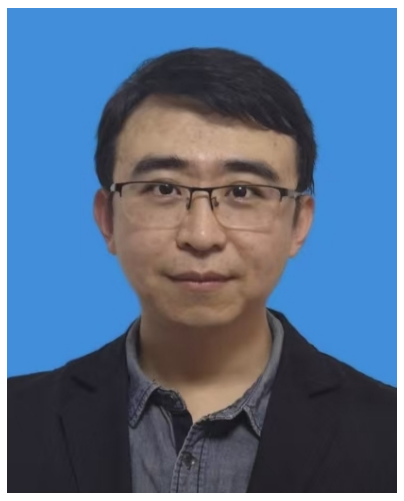

**Jiarun Song** (Member, IEEE) received the B.S. and Ph.D. degrees from Xidian University, Xi'an, China in 2009 and 2015, respectively. He is currently an Associate Professor with the State Key Laboratory of Integrated Services Networks, Xidian University. His research interests include QoE, visual quality assessment, XR interaction, and multimedia communication.

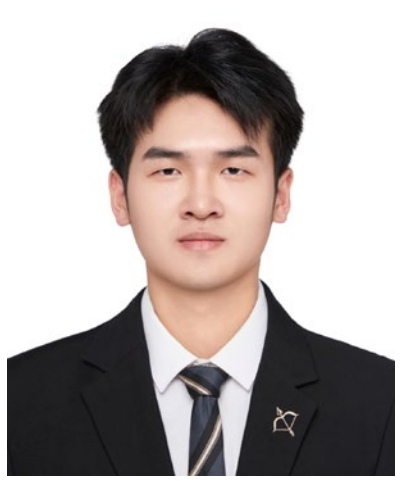

**Ninghao Wan** received the B.S. degree in Telecommunication Engineering from Xidian University, Xi'an, China, in 2024. He is currently a graduate student in the Xidian University. His research interests include Virtual/Augmented Reality, quality assessment, and multimedia communication.

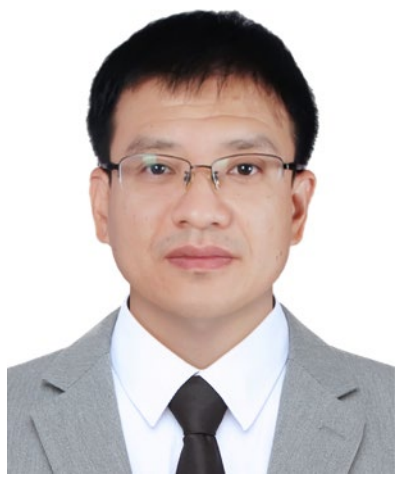

**Fuzheng Yang** (Member, IEEE) received the B.E. degree in Telecommunication Engineering, the M.E. degree and the Ph.D. in Communication and Information System from Xidian University, Xi'an, China, in 2000, 2003 and 2005, respectively. He became a lecturer and an Associate Professor in Xidian University in 2005 and 2006, respectively. He has been a professor of communications engineering with Xidian University since 2012. He is also an Adjunct Professor of School of Engineering in RMIT University. During 2006-2007, he served as a visiting scholar and postdoctoral researcher in Department of Electronic Engineering in Queen Mary, University of London. His research interests include video quality assessment, video coding and multimedia communication.

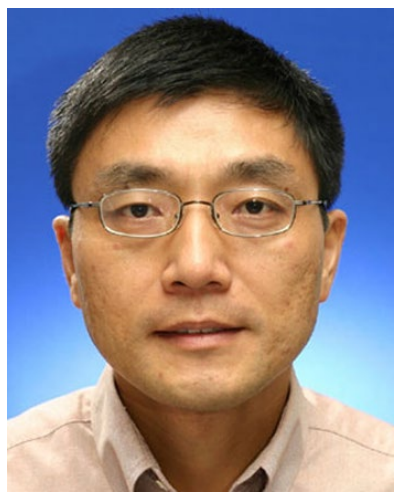

**Weisi Lin** (Fellow, IEEE) received the Ph.D. degree from the King's College, University of London, U.K. He is currently a Professor with the College of Computing and Data Science, Nanyang Technological University. His areas of expertise include image processing, perceptual signal modeling, video compression, and multimedia communication, in which he has published over 200 journal articles, over 230 conference papers, filed seven patents, and authored two books. He has been an invited/panelist/keynote/tutorial speaker at over 20 international conferences. He is a fellow of IET and an Honorary Fellow of Singapore Institute of Engineering Technologists. He has been the Technical Program Chair of IEEE ICME 2013, PCM 2012, QoMEX 2014, and IEEE VCIP 2017. He has been an Associate Editor of IEEE TRANSACTIONS ON IMAGE PROCESSING, IEEE TRANSACTIONS ON CIRCUITS AND SYSTEMS FOR VIDEO TECHNOLOGY, IEEE TRANSACTIONS ON MULTIMEDIA, and IEEE SIGNAL PROCESSING LETTERS. He was a Distinguished Lecturer of Asia-Pacific Signal and Information Processing Association (APSIPA) from 2012 to 2013 and the IEEE Circuits and Systems Society from 2016 to 2017.